\documentclass
[prl,twocolumn,superscriptaddress,floatfix,showpacs,10pt]{revtex4}%
\usepackage{amsfonts}
\usepackage{amsmath}
\usepackage{amssymb}
\usepackage{graphicx}%

\begin{document}

\title{On Background Independence and Duality Invariance in String Theory}

\newcommand{\be}{\begin{equation}}
\newcommand{\ee}{\end{equation}}
\newcommand{\ben}{\begin{displaymath}}
\newcommand{\een}{\end{displaymath}}
\newcommand{\bea}{\begin{eqnarray}}
\newcommand{\eea}{\end{eqnarray}}
\newcommand{\nn}{\nonumber}
\newcommand{\non}{\nonumber\\}
\newcommand{\bean}{\begin{eqnarray*}}
\newcommand{\eean}{\end{eqnarray*}}
\newcommand{\beqs}{\begin{eqnarray}}
\newcommand{\eeqs}{\end{eqnarray}}

\newcommand{\nin}[1] {\underline{\phantom{h}}\hskip-6pt {#1}}

\author{Olaf Hohm}
\email{ohohm@scgp.stonybrook.edu}
\affiliation{Simons Center for Geometry and Physics, 
Stony Brook University, 
Stony Brook, NY 11794-3636, USA}

\begin{titlepage}

\end{titlepage}

\begin{abstract}

Closed string theory exhibits an $O(D,D)$ duality symmetry on tori, which 
in double field theory is manifest before compactification.  
I prove that  to first order in  $\alpha'$ 
there is no manifestly background independent and duality invariant formulation of bosonic string theory 
in terms of metric, $b$-field and dilaton.   
To this end I use $O(D,D)$ invariant second order perturbation theory 
around flat space to show that the unique background independent 
candidate expression for the gauge algebra at order $\alpha'$ 
is inconsistent with the Jacobi identity.  
A background independent formulation 
exists instead for frame variables subject to 
$\alpha'$-deformed frame transformations 
(generalized 
Green-Schwarz transformations). 
Potential applications for 
curved backgrounds, 
as in cosmology, are discussed.

\end{abstract}

\maketitle

\setcounter{equation}{0}

String theory is a most promising 
candidate for a complete theory of quantum gravity. Remarkably, 
already \textit{classical} string theory generalizes Einstein's theory 
of general relativity 
by an infinite number of higher-derivative corrections,  
governed by the inverse string tension $\alpha'$. 
Moreover, new symmetries and dualities emerge together with 
new states. 
On toroidal backgrounds, string theory describes winding modes in addition 
to the massive Kaluza-Klein modes, 
which transform into each other under the T-duality 
group $O(D,D,\mathbb{Z})$. 
The spacetime theory for the massless fields, which truncates the 
Kaluza-Klein and winding modes but may include   
$\alpha'$ corrections, exhibits a continuous  $O(D,D,\mathbb{R})$
symmetry \cite{Meissner:1991zj}.

It is plausible that such characteristics are relevant in situations 
where general relativity is no longer applicable,  
as in the cosmology of the very early universe, and thereby lead to 
potentially testable phenomena. Since the 1980s various  
string cosmology proposals have been put forward that utilize 
these novel features \cite{Brandenberger:1988aj,Tseytlin:1991xk}. 
For instance, the proposal of \cite{Gasperini:2002bn} employs  
the $O(D,D)$ symmetry, together with the higher-derivative 
corrections, to argue that the big bang singularity is 
replaced by a smooth solution, possibly  already at the classical level. 
Without detailed control over these $\alpha'$ corrections, however, 
it is difficult to test such proposals. 
Given the recent renewed interest in 
higher-derivative gravity, particularly in cosmology and 
Starobinsky or $R^2$ inflation \cite{Planck:2013jfk}, it seems especially important  to have 
a formulation of string theory, at least classically,
 that is manifestly duality invariant and 
includes the $\alpha'$ corrections.

Although a complete off-shell 
formulation exists in the form of closed string field theory (CSFT),
it is difficult to use it for applications of the type just discussed. 
In particular, there is no \textit{background independent} 
formulation, which would be useful  in order to treat different curved backgrounds 
(as arising in cosmology) 
and their perturbations in one framework.  
In this letter I will employ CSFT \cite{Kugo:1992md} and double field theory (DFT)  
\cite{Siegel:1993th,Hull:2009mi,Hohm:2010jy,
Hohm:2013jaa,Hohm:2014eba} 
to prove an unexpected result: 
there is no manifestly background independent and manifestly 
duality invariant formulation of bosonic (or heterotic) string theory in terms of the universal massless fields 
(metric, $b$-field and dilaton)  when including the first $\alpha'$ correction. 
I will then show, however, that such a formulation can be obtained by means of a
frame or vielbein formalism, introducing extra fields as pure gauge modes under an  
$\alpha'$-deformed local frame symmetry, as proposed by Marques and Nunez \cite{Marques:2015vua}. 
These gauge transformations uniquely determine the first order $\alpha'$ corrections and 
generalize the Green-Schwarz 
transformations  
needed for anomaly cancellation \cite{Hohm:2014eba,Marques:2015vua}.  
In this letter, only the main results are given, while 
the technical
details 
will appear in  \cite{Hohmtoappear}.

While it has long been appreciated that a frame formulation often provides 
significant technical simplifications, it is usually not compulsory 
for a purely bosonic theory. [Although it has been suggested that frame variables 
may be needed for a non-perturbative formulation of 
quantum gravity \cite{Ashtekar:1986yd}.] 
 This changes when coupling fermions to gravity, which requires 
a frame formalism with a local Lorentz symmetry whose pure gauge modes
can only be gauged away when expanding around a background. 
The novelty discussed in this letter is that a purely bosonic theory requires a frame formulation.
In particular, this implies obstacles for any formulation of string theory that aims to make 
background independence and duality invariance manifest in terms of `metric-like' fields.

We begin by recalling  
the notion of background independence. A theory is manifestly background independent 
if it does not use a background structure, such as general relativity 
written in terms of the full metric tensor. Even if a theory does depend on a background,  
it may still be secretly background independent, as in general relativity  
with a metric $g$ expanded about a background $\bar{g}$: $g = \bar{g} + h$.  
Background independence is then verified  by 
showing that a shift of the background can be absorbed into an 
opposite shift of the fluctuation, 
\be\label{backgroundshifts}
 \delta_{\chi}\bar{g}  =  -\chi\,, \quad
 \delta_{\chi}h  =  \chi\,. 
\ee 
While there is no manifestly background independent formulation of 
string theory, CSFT is actually background independent 
in the above sense \cite{Sen:1993mh}.

Let us now consider string theory, using CSFT \cite{Kugo:1992md},  on a constant background 
$E_{ij}=G_{ij}+B_{ij}$, $i,j=1,\ldots, D$, encoding background metric and antisymmetric Kalb-Ramond field.  
The string field $|\Psi\rangle$ takes values in the Hilbert space of the first-quantized string and 
reads schematically 
 \be
  |\Psi\rangle  = \int {\rm d}P\, e_{ij}(P)  \alpha_{-1}^{\, i}\bar{\alpha}_{-1}^{\, j}c_1\bar{c}_1 |P\rangle +\cdots\,, 
 \ee 
where $\alpha_{-1}^{\,i}$ and $\bar{\alpha}_{-1}^{\, i}$ are the left- and right-moving oscillators 
of the world-sheet theory, and $c_1$ and $\bar{c}_1$ are ghosts. 
For our present purposes it is sufficient to note that 
the non-symmetric component field $e_{ij}$ encoding metric and $b$-field fluctuations 
can be thought of as carrying left- and right-handed indices, associated with the two oscillators, 
and that on toroidal backgrounds this field depends on doubled coordinates, 
corresponding to Kaluza-Klein and winding modes. 
String theory then exhibits a manifest $O(D)\times O(D)$ symmetry, 
which 
together with the $GL(D)$ diffeomorphism invariance and the invariance under 
shifts of the  $b$-field implies the $O(D,D)$ duality symmetry.

Let us review the $O(D,D)$ symmetry and the DFT 
formulation \cite{Siegel:1993th,Hull:2009mi,Hohm:2010jy}, which has recently 
been extended to higher order in $\alpha'$ 
\cite{Hohm:2013jaa,Hohm:2014eba,Marques:2015vua,Hohm:2015mka}.  
Writing the $O(D,D)$ group element and the invariant metric 
as
 \be
  h  \equiv  \begin{pmatrix}   a & b\\
  c & d \end{pmatrix}  \in  O(D,D)\,, \quad
  \eta_{MN} \equiv   \begin{pmatrix}   0 & {\bf 1}\\
  {\bf 1} & 0 \end{pmatrix}\,, 
 \ee
and the doubled coordinates as $X^M=(\tilde{x}_i,x^i)$, 
the field $e_{ij}$  transforms as 
 \be\label{eTRANS}
  e_{ij}'(X') = (M^{-1})_i{}^k (\bar{M}^{-1})_j{}^l e_{kl}(X)\;, 
 \ee  
where $X'=hX$, and $M= d^t-Ec^t$, $\bar{M}=d^t+E^t c^t$, 
which is accompanied by an appropriate transformation of the background.  
On toroidal backgrounds, the massless fields, together with their 
Kaluza-Klein and winding modes, provide a consistent subsector of the full string theory, as
recently proved by Sen \cite{Sen:2016qap}. While this theory, 
referred to as weakly constrained DFT, is not known explicitly beyond cubic order, 
a subsector of this theory, referred to as strongly constrained DFT, is known to all orders 
in a background independent form \cite{Hohm:2010jy}. 
This DFT is then applicable to arbitrary (geometric) string backgrounds and 
subject to the `strong constraint' 
 \be\label{SECTION}
  \eta^{MN}\partial_M\partial_NA  \equiv  \partial^M\partial_MA  =  0\;, \quad 
  \partial^MA\,\partial_MB  =  0\,, 
 \ee
for all fields $A, B$. This implies that the fields depend on only half of the coordinates, 
e.g.~$\tilde{\partial}^iA=0$ for all $A$. 

To zeroth order in $\alpha'$, a manifestly background independent 
formulation of DFT exists 
for  ${\cal E}_{ij}  \equiv  g_{ij} + b_{ij}$, 
which is related to the perturbative CSFT field variable $e_{ij}$ by the background expansion 
 \be\label{calEExpansionIntro}
  {\cal E}_{ij}  =  E_{ij} + e_{ij} +\tfrac{1}{2} e_{i}{}^{k} e_{kj}  +  {\cal O}(e^3)\,. 
 \ee
The $O(D,D)$ duality (\ref{eTRANS}) then becomes 
a genuine invariance, realized through the background independent 
fractional-linear transformations 
 \be\label{calEtransf}
  {\cal E}'(X')  =  (a{\cal E}(X)+b)(c{\cal E}(X)+d)^{-1}\;, 
 \ee   
corresponding to the non-linear realization based on $O(D,D)/O(D)\times O(D)$. 
The $O(D,D)$ symmetry can be linearized by introducing  
the generalized metric 
  \be\label{firstH}
  {\cal H}_{MN}  =   \begin{pmatrix}    g^{ij} & -g^{ik}b_{kj}\\[0.5ex]
  b_{ik}g^{kj} & g_{ij}-b_{ik}g^{kl}b_{lj}\end{pmatrix}\,,
 \ee
satisfying the constraint ${\cal H}\eta{\cal H}=\eta$. 
The transformation rule that is equivalent to (\ref{calEtransf}) reads 
 \be\label{HPRIME}
  {\cal H}_{MN}^{\prime}(X')  =  (h^{-1})^K{}_M\, (h^{-1})^L{}_N\, {\cal H}_{KL}(X)\,. 
 \ee 
Expanding about a  
background generalized metric $\bar{\cal H}_{MN}$, the fluctuations are constrained 
in order to preserve the constraint on ${\cal H}$: 
  \be\label{expandHAAAINntro}
  \begin{split}
  {\cal H}_{MN}   =   &\;\, \bar{\cal H}_{MN} + h_{\,\nin{M}\bar{N}}+ h_{\,\nin{N}\bar{M}}\\
   &\, -\tfrac{1}{2} h^{\,\nin{K}}{}_{\bar{M}} h_{\,\nin{K}\bar{N}} + \tfrac{1}{2} 
   h_{\,\nin{M}}{}^{\bar{K}}  h_{\,\nin{N}\bar{K}}   +   {\cal O}(h^3)\,, 
 \end{split}
 \ee 
where we introduced projected 
$O(D,D)$ indices, defined for a vector by $V_{\,\nin{M}}=P_M{}^{N}V_N$, $V_{\bar{M}}=\bar{P}_M{}^{N}V_N$, 
where 
\be\label{PROjectors}
  P  =   \tfrac{1}{2}({\bf 1} - \bar{\cal H})\,, \quad 
\bar{P} =  \tfrac{1}{2}({\bf 1} + \bar{\cal H})
\ee  
satisfy the projector relations $P^2=P, \bar{P}^2=\bar{P}, P\bar{P}=0$ as a consequence 
of the constraint $\bar{\cal H}\eta\bar{\cal H}=\eta$. 
The two fluctuation fields $e_{ij}$ and $h_{\,\nin{M}\bar{N}}$ utilized in (\ref{calEExpansionIntro}) 
and (\ref{expandHAAAINntro}) are essentially equivalent, c.f.~(\ref{ehoncemore}) below.

Let us briefly review the gauge symmetries to zeroth order in $\alpha'$, 
which act via a generalized Lie derivative: 
 \be\label{fullHGauge}
  \delta_{\xi}^{(0)}{\cal H}_{MN} = 
  \xi^K\partial_K{\cal H}_{MN} + K_M{}^K {\cal H}_{KN}+K_N{}^K {\cal H}_{MK}
 \ee
where $K_{MN}=2\partial_{[M}\xi_{N]}$ and indices are raised and lowered
with $\eta_{MN}$. Gauge parameters of the form $\xi^M=\partial^M\chi$ 
are trivial and do not transform fields. 
The gauge transformations close, $[\delta^{(0)}_{\xi_1},\delta^{(0)}_{\xi_2}]=\delta_{{F}^{(0)}(\xi_1,\xi_2)}$,  
according to the `C-bracket' 
 \be\label{CBRacket}
   {F}^{(0)}(\xi_1,\xi_2)^M 
   \equiv  
  2 \xi_{[2}^N\partial_N\xi_{1]}^M - \xi_{[2}^{N}\partial^M \xi_{1]N}  \,. 
 \ee

Let us now consider the first $\alpha'$ correction. The cubic theory 
around constant backgrounds was derived from bosonic CSFT in \cite{Hohm:2014eba}. 
The gauge transformations linear in fields receive $\alpha'$ corrections, 
corresponding to the following deformation of the gauge algebra 
\begin{equation*}
\begin{split}
{F}^{(1)}(\xi_1,\xi_2)^M  =  -\tfrac{a}{2} K_{[2}^{\,\nin{K}\,\nin{L}} \, \partial^M K_{1]\,\nin{K}\,\nin{L}}
   +\tfrac{b}{2}   K_{[2}^{\bar{K}\bar{L}} \partial^M K_{1] \bar{K}\bar{L}}\,, 
 \end{split}  
\end{equation*}
where $K_{1,2 MN}=2\partial_{[M}\xi_{1,2 N]}$ and explicit factors of $\alpha'$ are suppressed. 
For $a=b=1$ this corresponds to bosonic string theory, for $a=1, b=0$ to 
heterotic string theory and for $a=1, b=-1$ to the theory in \cite{Hohm:2013jaa}.  
To this order, the algebra is field independent but background dependent. 
Using the Noether procedure, the gauge structure can be extended uniquely to second order 
in $h$ and still first order in $\alpha'$. 
The field $h$ then enters the gauge transformations quadratically, and the gauge algebra 
is field dependent, linear in $h$.  

We now ask whether there is a manifestly background independent gauge algebra 
in terms of the full generalized metric ${\cal H}_{MN}$,  
$[\delta_{\xi_1},\delta_{\xi_2}]=\delta_{F(\xi_1,\xi_2;{\cal H})}$, that reproduces this result to 
second order upon expanding as in (\ref{expandHAAAINntro}). 
There is a unique such expression in terms of the generalized metric, 
which reads for the bosonic string, $a=b=1$, 
 \be\label{UNiqueF}
  \begin{split}
   F^{(1)}(\xi_1,\xi_2;{\cal H})^M  &= 
   \tfrac{1}{2}{\cal H}^{KL} K_{[2 K}{}^{P}\partial^M K_{1]LP}\\[0.5ex]
   &\hspace{-3.3ex} -  \tfrac{1}{4}  {\cal H}^{K}{}_{R}\,\partial^M {\cal H}^{RL} {\cal H}^{PQ} K_{[2KP} K_{1]LQ} \;, 
  \end{split}
 \ee  
while $F^{(0)}(\xi_1,\xi_2)$ is given in (\ref{CBRacket}). 
One cannot write higher order terms in ${\cal H}$, because by the constraint on the generalized metric 
they can be reduced to terms with fewer fields. Hence, this expression is unique. 
We will now prove, however, that it does not define a consistent gauge algebra. 
A necessary condition is the Jacobi identity $\sum_{\rm cycl.} [[\delta_{\xi_1},\delta_{\xi_2}],\delta_{\xi_3}]=0$. 
One may verify that to first order in $\alpha'$ this implies 
 \be
 \begin{split}
  \sum_{\rm cycl}&\;\Big\{\delta_{\xi_1}^{(0)}F^{(1)}(\xi_2,\xi_3;{\cal H})+
  F^{(0)}(\xi_1,F^{(1)}(\xi_2,\xi_3;{\cal H}))\\[-1.7ex]
 &\;\; +F^{(1)}(\xi_1,F^{(0)}(\xi_2,\xi_3);{\cal H})\Big\} = {\rm trivial}\;, 
 \end{split}
 \ee 
with the lowest order gauge transformations $\delta^{(0)}$ in (\ref{fullHGauge}). 
This needs to be a trivial gauge parameter $\xi^M=\partial^M\chi$. 
A direct computation with (\ref{fullHGauge}) and (\ref{UNiqueF}) shows, however, that 
this condition is not satisfied. Thus, there is 
no background independent formulation in terms of the generalized metric that is compatible 
with the perturbative results from bosonic CSFT.

Instead, I will now turn to a frame formalism that slightly generalizes \cite{Marques:2015vua}
and allows for a 
consistent ${\cal O}(\alpha')$ deformation of the local frame transformations, 
which in turn provides a background independent formulation that is consistent  with 
the perturbative results. The background independent frame field is 
denoted by $E_{A}{}^{M}$, where $A=(a,\bar{a})$ are flat indices 
w.r.t.~$GL(D)\times GL(D)$. The frame field is subject to the constraint that the `flattened' 
$O(D,D)$ metric is block-diagonal:
 \be\label{TANGENTmetric}
  {\cal G}_{AB} \equiv  E_{A}{}^{M} E_{B}{}^{N} \eta_{MN}   =      \begin{pmatrix}    {\cal G}_{ab} & 0\\
  0 & {\cal G}_{\bar{a}\bar{b}} \end{pmatrix} . 
 \ee 
The local $GL(D)\times GL(D)$ frame transformations read 
 \be\label{localFRAME}
  \delta_{\Lambda}E_{A}{}^{M}  =  \Lambda_A{}^{B} E_{B}{}^{M}\;, 
 \ee
 but they are $\alpha'$-deformed in that the transformation matrix is given by 
  \be\label{DeformedMAtrix}
  \Lambda_A{}^{B}  =   \begin{pmatrix}    \Lambda_a{}^b &  \Sigma_a{}^{\bar{b}}(\Lambda,E)\\[0.3ex]
    \Sigma_{\bar{a}}{}^{b}(\Lambda,E) & \Lambda_{\bar{a}}{}^{\bar{b}} \end{pmatrix}    \;, 
 \ee
with $\Sigma$ being defined in terms of derivatives of the gauge parameters $\Lambda_a{}^b$ and 
${\Lambda}_{\bar{a}}{}^{\bar{b}}$, 
 \be
  \Sigma_a{}^{\bar{b}}  \equiv 
   \tfrac{a}{2}  {\cal D}_{a}\Lambda_c{}^{d}\,\omega^{\bar{b}}{}_{d}{}^{c}
   +\tfrac{b}{2}  {\cal D}^{\bar{b}}\Lambda_{\bar{c}}{}^{\bar{d}}\,\omega_{a\bar{d}}{}^{\bar{c}}\;. 
 \ee 
Here ${\cal D}_{A}  \equiv  E_{A}{}^{M}\partial_M$, and $\omega_{\bar{a}b}{}^{c}$
and $\omega_{a\bar{b}}{}^{\bar{c}}$ are generalized spin connections \cite{Siegel:1993th}. 
These are of first order in derivatives and hence the $\Sigma$ terms carry two derivatives and are 
of order $\alpha'$. 
Remarkably, these gauge transformations  close to first order in $\alpha'$, 
with an $\alpha'$ correction 
of the diffeomorphism algebra 
 \be\label{xiDEFor}
  \xi_{12}^{(1)M}  =  \tfrac{a}{2} \Lambda_{[2c}{}^{d}\, \partial^M\Lambda_{1]d}{}^{c}
  -\tfrac{b}{2} \Lambda_{[2\bar{c}}{}^{\bar{d}}\, \partial^M\Lambda_{1]\bar{d}}{}^{\bar{c}} \;, 
 \ee
and an $\alpha'$ correction of the $GL(D)\times GL(D)$ algebra  
  \be\label{Lambda12prime}
  \begin{split}
   \Lambda^{(1)}_{12a}{}^{b}  &=  a\, {\cal D}_a\Lambda_{[2c}{}^{d}\,{\cal D}^b\Lambda_{1]d}{}^{c}
   -b\, {\cal D}_{a}\Lambda_{[2\bar{c}}{}^{\bar{d}}\,{\cal D}^b\Lambda_{1]\bar{d}}{}^{\bar{c}} \,,  \\
    \Lambda^{(1)}_{12\bar{a}}{}^{\bar{b}} &=  a\, {\cal D}_{\bar{a}}\Lambda_{[2c}{}^{d}\,
    {\cal D}^{\bar{b}}\Lambda_{1]d}{}^{c}
   -b\,{\cal D}_{\bar{a}}\Lambda_{[2\bar{c}}{}^{\bar{d}}\,{\cal D}^{\bar{b}}\Lambda_{1]\bar{d}}{}^{\bar{c}} \,. 
  \end{split} 
  \ee

Let us now expand about a background $\bar{E}_{A}{}^{M}$,  
  \be\label{backgroundExp}
  E_{A}{}^{M}  =  \bar{E}_{A}{}^{M} - h_{A}{}^{B}\bar{E}_{B}{}^{M}\;, 
 \ee 
with a fluctuation field carrying flat indices as in 
\cite{Siegel:1993th,Hohm:2011dz}. To lowest order in $\alpha'$ and in the 
number of fields, $GL(D)\times GL(D)$ acts as a St\"uckelberg symmetry, 
$\delta h_{ab}=\Lambda_{ab}$, $\delta h_{\bar{a}\bar{b}}=\Lambda_{\bar{a}\bar{b}}$. 
Thus, we can fix a gauge by setting $h_{ab}=h_{\bar{a}\bar{b}}=0$, 
after which 
the constraint (\ref{TANGENTmetric}) implies $h_{a\bar{b}}=-h_{\bar{b}a}$, 
which encodes the physical field. It is related to the perturbative variables above by \cite{Hohm:2011dz,Hohm:2014eba}
 \be\label{ehoncemore}
  h_{a\bar{b}}  =  \bar{E}_{a}{}^{i} \bar{E}_{\bar{b}}{}^{j} e_{ij} 
  =  \tfrac{1}{2} {\bar E}_{a}{}^{M}{\bar E}_{\bar{b}}{}^{N}h_{\,\nin{M}\bar{N}} \,.
 \ee
The gauge fixing condition 
requires compensating frame transformations, which in turn lead to deformed 
generalized diffeomorphisms with parameter $\xi^M$. 
The resulting gauge transformations to second order in fields, and the gauge algebra to 
first order in fields, agree precisely, up to $O(D,D)$ covariant field and parameter redefinitions,  
with those found by the Noether procedure from CSFT. This confirms that the $\alpha'$-deformed 
frame formalism provides the proper background independent formulation of the first order $\alpha'$
corrections of string theory. 

Starting from the frame formulation, 
we can now give an alternative proof that for general $a, b$ there is no generalized metric formulation, 
using the notion of background independence expressed in (\ref{backgroundshifts}). 
We first note that the expansion (\ref{backgroundExp}) is invariant under 
background shifts $\delta_{\Delta} \bar{E}_{A}{}^{M}=\Delta_{A}{}^{B}\bar{E}_{B}{}^{M}$, provided they are accompanied 
by the field transformation 
 \be\label{Backgroundshifts}
  \delta_{\Delta} h_{A}{}^{B} =  \Delta_{A}{}^{B}-h_{A}{}^{C}\Delta_{C}{}^{B}\,. 
 \ee
Imposing the gauge condition $h_{ab}=h_{\bar{a}\bar{b}}=0$, these transformations receive 
$\alpha'$ corrections through compensating frame transformations. 
Writing these in terms of $h_{\,\nin{M}\bar{N}}$ via (\ref{ehoncemore}) 
and setting $\chi_{\,\nin{M}\bar{N}}=2\bar{E}_{M}{}^{a}\bar{E}_{N}{}^{\bar{b}}\Delta_{a\bar{b}}$, 
one obtains, up to $O(D,D)$ covariant field redefinitions, 
 \be\label{APLphaPRimeDEformed}
 \begin{split}
  \delta_{\chi}h_{\,\nin{M}\bar{N}}   =  &\, 
    \chi_{\,\nin{M}\bar{N}} + \tfrac{1}{2} \chi^{\,\nin{K}}{}_{\bar{M}} h_{\,\nin{K}\bar{N}}
   -\tfrac{1}{2} \chi_{\,\nin{N}}{}^{\bar{K}}\,h_{\,\nin{M}\bar{K}}
   +\cdots \\
   &\, +\tfrac{1}{16}(a+b)\partial_{\,\nin{M}}h_{\,\nin{K}\bar{L}}\chi^{\,\nin{P}\bar{L}} h^{\,\nin{K}\bar{Q}} 
   \partial_{\bar{N}}h_{\,\nin{P}\bar{Q}}\,, 
  \end{split}
 \ee  
where the ellipsis denote higher order terms in $h$ without derivatives. 
For a $\bar{\delta}_{\chi}h$ defined by the terms in the first line only 
one would have background independence in terms of a constrained 
generalized metric in that (\ref{expandHAAAINntro}) satisfies 
$\bar{\delta}_{\chi}{\cal H}_{MN}=0$ for background transformations 
$\bar{\delta}_{\chi}\bar{\cal H}_{MN}=-\chi_{\,\nin{M}\bar{N}}-\chi_{\,\nin{N}\bar{M}}$.
However, (\ref{APLphaPRimeDEformed}) also contains the higher-derivative 
term in the second line, which is not removable by a field  redefinition. 
Therefore, a generalized metric formulation does not exist to first order in $\alpha'$, 
unless $a=-b$. A manifestly duality invariant 
formulation exists as a frame formalism with pure gauge degrees of freedom
that can be eliminated in a duality covariant way only upon expanding around a background.

In order to elucidate this point, let us fix the $GL(D)\times GL(D)$ gauge symmetry 
before expanding about a background as follows
 \be\label{gaugeFIXEDframe}
  E_{A}{}^{M}  =   \begin{pmatrix}    E_{ai} & E_{a}{}^{i}  \\[0.3ex]
  E_{\bar{a}i} & E_{\bar{a}}{}^{i} \end{pmatrix} 
  =   \begin{pmatrix}    -{\cal E}_{ai} & \delta_{a}{}^{i}  \\[0.3ex]
  {\cal E}_{i\bar{a}} & \delta_{\bar{a}}{}^{i} \end{pmatrix} , 
 \ee  
which satisfies (\ref{TANGENTmetric}) and is written in terms of a non-symmetric metric 
${\cal E}_{ij}$, where the Kronecker deltas are used to identify flat and curved indices.  
The gauge fixing requires compensating frame transformations, however, which leads  
to deformed $O(D,D)$ transformations. 
Thus, to first order in $\alpha'$ the field ${\cal E}_{ij}$ cannot be identified with (\ref{calEExpansionIntro}) 
transforming as (\ref{calEtransf}).  
The gauge algebra of generalized diffeomorphisms is also deformed. 
It can be written in terms of the components of $K_{MN}=2\partial_{[M}\xi_{N]}$, 
up to parameter redefinitions, as
\begin{equation*}
 \begin{split}
  &\xi_{12}^{(1)M}  =    \tfrac{a-b}{2}\,\partial_K\xi_{[2}^L\, \partial^M\partial_L\xi_{1]}^K\\
  &-\tfrac{a+b}{2}\Big\{({\cal E}_{kl}+{\cal E}_{lk})K_{[2}^{kp} \partial^M K_{1]p}{}^{l} 
  -{\cal E}_{kp}\,\partial^M{\cal E}_{ql}\,K_{[2}^{kl} K_{1]}^{pq}\Big\}. 
 \end{split}
 \end{equation*}
The term in the first line, which survives for $a=-b$, is manifestly $O(D,D)$ covariant
and corresponds to the deformation found in \cite{Hohm:2013jaa}, 
for which indeed there is a generalized metric formulation \cite{Hohm:2015mka}. 
This gauge algebra is non-zero for $\tilde{\partial}^i=0$;  
it encodes, in particular, the Green-Schwarz deformation \cite{Hohm:2014eba}. 
In contrast, the terms in the second line are not covariant under the undeformed 
$O(D,D)$ and vanish for $\tilde{\partial}^i=0$, 
and so on the physical subspace the diffeomorphism algebra is not deformed.

I close with some general remarks and an outlook. 
It would be important to extend the frame formalism 
beyond first  order in $\alpha'$, which almost certainly will require  
further terms in the gauge transformations and/or the gauge algebra.   
It is conceivable that there is an exact formulation,  
as for the invariant subsector in \cite{Hohm:2013jaa}. 
Moreover, the results here could be useful for string cosmology. 
The dynamics of Friedmann-Robertson-Walker (FRW) backgrounds with 
metric ${\rm d}s^2=-{\rm d}t^2+a^2(t){\rm d}{\bf x}^2$
is effectively governed by a   
one-dimensional theory, for which the fields do not depend on 
$\bf{x}$. Hence, there is no obstacle for making the $O(d,d)$ acting on the spatial directions 
manifest in a generalized metric formulation,  
 in agreement with ref.~\cite{Meissner:1996sa}, because 
${\cal O}(\alpha')$ terms as in (\ref{APLphaPRimeDEformed}) vanish. 
However, in cosmological perturbation theory one considers 
fluctuations $h(t,{\bf x})$ around FRW, whose spatial derivatives no longer 
vanish so that 
the $\alpha'$-deformed geometry discussed here becomes important.

\noindent\textbf{Acknowledgements}:  
I would like to thank Diego Marques and Warren Siegel for discussion. 
I am especially 
indebted to Ashoke Sen and Barton Zwiebach for early collaborations and extensive discussions 
on these and related questions. 
This work is supported by a DFG Heisenberg Fellowship of the German Science Foundation (DFG).

\end{document}